\documentclass[12pt]{iopart}
\usepackage{epsf,iopams,graphicx,latexsym}
\def\rmd{{\rm d}}
\def\ApJ{{\em Astrophys. J. }}
\def\AJ{{\em Astron. J. }}
\def\MNRAS{{\em Mon. Not. Roy. Astron. Soc. }}
\def\lsim{\mathrel{\rlap{\lower4pt\hbox{\hskip1pt$\sim$}}
    \raise1pt\hbox{$<$}}}
\newcommand{\erf}[1]{(\ref{#1})}

\begin{document}

\title[Event rates for LISA EMRI sources]{Event rate estimates for LISA extreme mass ratio capture sources}
\author{Jonathan R Gair$^\dag$, Leor Barack$^\ddag$, Teviet Creighton$^\S$, Curt Cutler$^\|$, Shane L Larson$^\P$, E Sterl Phinney$^\dag$ and Michele Vallisneri$^{+}$}
\address{$^\dag$Theoretical Astrophysics, California Institute of 
Technology, Pasadena, CA 91125}
\address{$^\ddag$Department of Physics and Astronomy and Center for 
Gravitational Wave Astronomy, University of Texas at Brownsville, 
Brownsville, TX 78520}
\address{$^\S$LIGO Laboratory, California Institute of Technology, 
Pasadena, CA 91125}
\address{$^\|$Max-Planck-Institut fuer Gravitationsphysik, 
Albert-Einstein-Institut, Am Muchlenberg 1, D-14476 Golm bei Potsdam, 
Germany}
\address{$^\P$Space Radiation Laboratory, California Institute of 
Technology, Pasadena, CA 91125}
\address{$^{+}$Jet Propulsion Laboratory, California Institute of 
Technology, Pasadena, CA 91109}
\ead{jgair@tapir.caltech.edu}

\date{\today}

\begin{abstract}
One of the most exciting prospects for the LISA gravitational wave observatory is the detection of gravitational radiation from the inspiral of a compact object into a supermassive black hole. The large inspiral parameter space and low amplitude of the signal makes detection of these sources computationally challenging. We outline here a first cut data analysis scheme that assumes realistic computational resources. In the context of this scheme, we estimate the signal-to-noise ratio that a source requires to pass our thresholds and be detected. Combining this with an estimate of the population of sources in the Universe, we estimate the number of inspiral events that LISA could detect. The preliminary results are very encouraging --- with the baseline design, LISA can see inspirals out to a redshift $z=1$ and should detect over a thousand events during the mission lifetime.
\end{abstract}
%`Omne ignotum pro magnifico est.' - Tacitus, {\em De Vita et Moribus Iulii Agricolae}

 \submitto{\CQG}

\maketitle

\section{Introduction}
The inspiral of a compact object (CO) into a supermassive black hole (SMBH) is an exciting potential source for LISA \cite{LISAppa}. The extreme mass ratio of these systems ensures that the CO acts as a probe of the gravitational potential of the SMBH. For the last several years before plunge, the orbit remains close to the SMBH and the emitted gravitational radiation effectively maps out this strong field region of the spacetime \cite{ryan97}. If the central object is really a Kerr black hole, extreme relativistic perihelion and Lense-Thirring precessions are evident in the zoom and whirl waveforms \cite{gk2002}. These complex features allow measurement of the mass and spin of the SMBH to unprecedented precision \cite{AK03}. If the central object is something other than a black hole, for instance a massive boson star, the difference may be evident in the gravitational waves. Extreme mass ratio inspiral (EMRI) waveforms thus provide a strong field test of GR and black hole physics. Methods are presently being developed to detect and characterize deviations from the Kerr predictions (e.g., \cite{collins04}), but here we will focus on detection of EMRIs into Kerr black holes.

The desire for LISA to detect a significant number of EMRIs during the mission has been driving the final LISA mission specification. A typical expected EMRI source will be buried in the detector noise. These signals can be extracted using matched filtering, but the total number of detections will depend on the frequency range and level of the LISA noise floor. We describe here a first cut effort to estimate the LISA detection rate, assuming a plausible data analysis technique that employs realistic computational resources. These preliminary results indicate that with its baseline design LISA should see about a thousand EMRI events during its lifetime. There appears to be no pressing need to modify the satellite design in order to enhance the EMRI event rate. More details of this work can be found in a white paper prepared for the LIST (LISA International Science Team) \cite{LIST03}.

In section~\ref{astro} we describe some astrophysical aspects of EMRIs, including the source parameters of a typical event and an estimate of the capture event rate. In section~\ref{detect} we outline a plausible detection scheme and then in section~\ref{rate} we estimate the number of events that LISA would see using this scheme, given the astrophysical rate estimate. We finish in section~\ref{discuss} with a brief discussion of these results and outline some of the remaining uncertainties.

\section{Astrophysics of EMRIs}
\label{astro}
\subsection{Source parameters}
EMRIs occur as a consequence of large angle scattering encounters between objects in the cusp of stars surrounding a supermassive black hole (SMBH) at the center of a galaxy. Such encounters may put a star onto an orbit that passes sufficiently close to the central black hole that gravitational radiation dominates the subsequent evolution and it becomes bound to the SMBH. The stellar orbit decays over time, due to the loss of energy and angular momentum in bursts of gravitational radiation emitted near periapse. Initially these bursts are widely separated in time and the radiation will not be detectable, but in the last few years of inspiral the source will be radiating continuously at frequencies to which LISA is sensitive. The typical frequency of the gravitational radiation is determined by the mass of the primary, $M$. The floor of the LISA noise curve ($\sim 0.003 {\rm Hz} \-- 0.03 {\rm Hz}$) therefore sets the mass range to which we are most sensitive at $M\sim 10^5M_{\odot} \-- 5\times10^6M_{\odot}$. LISA could detect nearby sources with other primary masses, but we concentrate on this mass range since it should dominate the event rate. The secondary has to be a compact body to avoid tidal disruption before plunge, so we consider white dwarfs ($m\sim0.6M_{\odot}$), neutron stars ($m\sim1.4M_{\odot}$) and both stellar mass ($m\sim10M_{\odot}$) and intermediate mass black holes ($m\sim100M_{\odot}$). The spin of the primary can take any value in the range $S/M^{2} = 0 \-- 1$, with spins around $0.9$ probably being typical, since this is approximately the point of spin equilibrium in black hole accretion models \cite{gamm04,mckinn04}. EMRI orbits are generally moderately eccentric at plunge, $e \sim 0 \-- 0.4$ depending on the periapse at capture, and can have any initial inclination to the black hole spin axis. The orbital periapse can take any value between the plunge periapse at a few $M$ and the capture periapse at $10$'s of $M$, but we restrict this range by only searching for inspirals that are in the last few years before plunge.

\subsection{Capture event rate}
There are two ingredients that go into an estimate of the frequency with which EMRIs occur in the Universe. The first is an estimate of the space density of supermassive black holes in the appropriate mass range, and the second is an estimate of the rate at which each black hole is consuming compact objects. The space density of SMBHs is constrained observationally, with the tightest measurement of black hole mass, $M_{\bullet}$, coming from the correlation with spheroid velocity dispersion, $\sigma$
\begin{equation}
\label{msigma}
M_{\bullet}=M_{\bullet,*}\,\,\left(\frac{\sigma}{\sigma_{*}}\right)^{\lambda}
\end{equation}
where $\lambda$, $M_{\bullet,*}$ and $\sigma_{*}$ are constants. We use $\sigma_{*}=90\,{\rm km\,s^{-1}}$ and constrain $\lambda$ and $M_{\bullet,*}$ from observations. 
%and we adopt $\lambda=5$ and $M_{\bullet,*}=3\times 10^6 \,h^{-1}_{65}\,M_{\odot}$, with the usual notation $h_{65} = H_{0}/65 
%{\rm km\,s^{-1}\,Mpc^{-1}}$. 
Merritt and Ferrarese \cite{mf01} find $\lambda =4.72$ and $M_{\bullet,*}=3 \times 10^6 \,M_{\odot}$. Tremaine \etal \cite{trem02} find $\lambda=4.02$ and $M_{\bullet,*}= 5 \times 10^6 \,M_{\odot}$, but use a nonstandard definition of dispersion. The galaxy velocity dispersion function may be constrained indirectly using galaxy luminosity functions and the $L\--\sigma$ correlation \cite{aller02}. In conjunction with the $M\--\sigma$ relation \erf{msigma}, this leads to a black hole mass function of the form
\begin{equation}
\label{bhdist}
M_{\bullet}\,\frac{\rmd N}{\rmd M_{\bullet}} = \phi_{*}\,\frac{\epsilon}{\Gamma\left(\frac{\gamma}{\epsilon}\right)} \, \left(\frac{M_{\bullet}}{M_{\bullet,*}}\right)^{\gamma}\, \exp\left( -\left(\frac{M_{\bullet}}{M_{\bullet,*}}\right)^{\epsilon}\right)
\end{equation}
where $\epsilon=3.08/\lambda$, $\Gamma(z)$ is the Gamma function and $\phi_{*}$ is a constant equal to the total number density of galaxies. Aller and Richstone \cite{aller02} use the value $\lambda=4.02$ from Tremaine \etal \cite{trem02} to set $\epsilon=0.8$ and constrain the parameters $\phi_{*}$, $M_{\bullet,*}$ and $\gamma$ according to galaxy type. These values are listed in Table~\ref{bhdistparams}. For the mass range of interest in this analysis, $M_{\bullet} \lsim 5 \times 10^6 M_{\odot}$, the total Aller and Richstone space density of black holes is approximately
\begin{equation}
M_{\bullet}\,\frac{\rmd N}{\rmd M_{\bullet}} = 3\times 10^{-3} \,h_{65}^{2}\,{\rm Mpc}^{-3}
\end{equation}
where $h_{65}=H_{0}/65 \,{\rm km\,s^{-1}\,Mpc^{-1}}$ is the Hubble parameter. Some Sc$\--$Sd galaxies have central black hole masses much lower than would be derived from their luminosities \cite{Scobs}. If we remove these from the sample, the black hole space density is reduced by a factor of two. We adopt this lowered value as the reference density for estimating the capture rate. In an interval $\Delta \log_{10}(M_{\bullet}/M_{\odot}) = 0.5$, the total space density of black holes is then $1.7 \times10^{-3} h_{65}^{2}{\rm Mpc}^{-3}$.

To estimate the rate at which each SMBH is capturing compact objects, we make use of Marc Freitag's simulation of the Milky Way \cite{freitag01}. That simulation used $\sigma=\sigma_{*}$ and $M_{\bullet}=4\times 10^6 \,M_{\odot}$ and predicts present day capture rates of $5\times10^{-6}\,{\rm y^{-1}}$, $10^{-6}\,{\rm y^{-1}}$ and $10^{-6}\,{\rm y^{-1}}$ for $0.6\,M_{\odot}$ white dwarfs, $1.4\,M_{\odot}$ neutron stars and $9\,M_{\odot}$ black holes respectively. Freitag does not simulate intermediate mass black hole remnants, but Madau and Rees \cite{madau01} estimate a dynamical friction rate of $2\times 10^{-9}\,{\rm y^{-1}}$ in the Milky Way for these objects. We assume optimistically that half of these are captured by gravitational radiation (the others being direct plunges) and use a capture rate of $1\times 10^{-9}\,{\rm y^{-1}}$. To extrapolate from these results to other central black hole masses we note that since these bodies are captured by large angle scattering \cite{sig97}, the rate of gravitational capture is comparable to the rate of direct plunges and each is approximately half of the dynamical friction rate. For stars with mass $m$ much greater than the mean, the dynamical friction timescale is 
\begin{equation}
\label{tdf}
t_{{\rm df}}\approx 0.3\,\frac{\sigma^{3}}{G^{2}\,m\,\rho_{*}\,\ln\Lambda}
\end{equation}
where $\rho_{*}$ is the local stellar density, $\sigma$ is the spheroid velocity dispersion and $\ln \Lambda \sim 5$ measures the range of impact parameters for stellar encounters in the cluster \cite{binney88}. Galaxies with SMBHs in the mass range to which LISA is sensitive tend to have isothermal density profiles in the core, $\rho_{*}=\sigma^{2}/(2\,\pi\,G\,r^{2})$. Substituting this profile and the relation \erf{msigma} with $\lambda=4$ into equation \erf{tdf}, we deduce, for each stellar component $m$, the radius, $r_{{\rm df}}(m)$, within which $t_{{\rm df}} < t$, as a function of time $t$. Assuming the cluster has an age $t\sim 10^{10}\,{\rm y}$ and all the stars within $r_{{\rm df}}(m)$ have sunk to the center, we find that the total mass in such stars scales as $M_{{\rm df}} (m) /M_{\bullet,*}\approx F\,(m/M_{\odot})^{1/2}\, (M_{\bullet}/M_{\bullet,*})^{3/8}$, where $F$ is the fraction of the total stellar mass in that component and $M_{\bullet,*}\sim 3\times 10^6 M_{\odot}$. Taking the mass in gravitational captures to be half this, we estimate the extreme mass ratio capture rate in a galaxy today to be
\begin{equation}
\label{captrate} \fl
\frac{1}{2}\,\frac{M_{{\rm df}} (m)}{m\,t} \approx \frac{F\,M_{\bullet,*}}{2\,m\,t}\,\left(\frac{M_{\bullet}}{M_{\bullet,*}}\right)^{\frac{3}{8}}\, \left(\frac{m}{M_{\odot}}\right)^{\frac{1}{2}} \approx 10^{-4}\,F\,\left(\frac{M_{\bullet}}{M_{\bullet,*}}\right)^{\frac{3}{8}}\, \left(\frac{m}{M_{\odot}}\right)^{-\frac{1}{2}}\,{\rm y^{-1}}
\end{equation}
where we again used $t \sim 10^{10}$y. Using a Kroupa IMF and standard initial-final mass relations, the mass fraction $F=0.2$ for $0.7\,M_{\odot}$ white dwarfs, $F=0.03$ for $10\,M_{\odot}$ black holes and $F=4\times 10^{-5}$ for $\sim100\,M_{\odot}$ Pop III black holes. Taking $M_{\bullet}=M_{\bullet,*}$, the rates predicted by equation \erf{captrate} agree with Freitag's simulation for $10\,M_{\odot}$ black holes and within a factor of two of the Madau and Rees estimate for $100\,M_{\odot}$ black holes. Equation \erf{captrate} over predicts Freitag's white dwarf rate by a factor of four, but the previous assumptions are not valid for white dwarfs since the capture time is longer than the Hubble time and mass segregation discriminates against these low mass stars. However, the $M_{\bullet}^{3/8}$ scaling with black hole mass should be fairly good for all stellar components and we use this to scale Freitag's rates for the Milky Way to other galaxy masses. Combining this with the black hole space density, we estimate the rate of mergers today for three ranges of SMBH mass, and four types of compact object capture. These rates are summarized in Table~\ref{ratetab}.

There are uncertainties in both the space density of black holes and the capture rate. The distribution of galaxy velocity dispersions can be constrained by direct observation, rather than using the indirect correlation with luminosity employed here. Sheth \etal \cite{sheth03} use SDSS data to measure velocity dispersions, and find the total SMBH space density in our range of interest to be an order of magnitude lower. However, the SDSS spectra do not have sufficient resolution to measure the dispersion in this black hole mass range, so this extrapolation should not be trusted. Hils and Bender \cite{hils95} estimate the white dwarf capture rate to be a factor of $150$ smaller than our extrapolation from Freitag, but they assume an adiabatic central density profile and only half the number of white dwarfs that modern IMFs predict. Sigurdsson and Rees \cite{sig97} predict a rate that is a factor of $50$ lower than Freitag's, but their central cusps were not fully self-consistent and they ignored mass segregation. In more recent simulations, Freitag \cite{freitag03} uses a new model and also finds capture rates for all species that are an order of magnitude lower. The new model uses a different IMF and cluster model and contains a fixed mass central SMBH while the old model grew the SMBH adiabatically from a tiny seed. We use the old results because they are consistent with the expression \erf{captrate} used to extrapolate to other central SMBH masses, but must allow for these uncertainties. A conservative rate for the white dwarfs would be $10^{-2}$ of those in Table~\ref{ratetab}. The black hole rates are more robust to stellar dynamics, but depend on the mass fraction of stellar mass black holes. While the increasing number of observed galactic black hole binaries give some confidence that this mass fraction is not too different from our assumptions, we should allow an order of magnitude or more uncertainty in the black hole rates.

%cut out Merritt and Ferrarese and Tremaine references?
%cut out bhdist equation?
%cut out bhdistparams table?
%cut out details of capture rate?
%cut out discussion of uncertainties?

%arguments for our event rate

\begin{table}
\begin{tabular}{l|l|l|l}
\hline &$M_{\bullet,*}$&$\phi_{*}$&$\gamma$ \\
Galaxy Type&$(10^7\,h_{65}^{-1}\,M_{\odot})$&$(10^{-3}\,h_{65}^{3}\,Mpc^{-3})$& \\ \hline
E&$17$&$2.3$&$0.12$\\ \hline S0&$5$&$33.7$&$0.046$ \\ \hline Sa$\--$Sb&$2$&$5.0$&$0.32$ \\ \hline Sc$\--$Sd&$0.5$&$29.4$&$0.03$ \\ \hline
\end{tabular}
\caption{Parameters for black hole space densities in equation \erf{bhdist}.}
\label{bhdistparams}
\end{table}

\begin{table}
\begin{tabular}{l|l|l|l|l}
\hline
 $M_{\bullet}$ & \multicolumn{4}{|c}{Merger rate $\mathcal{R}$}
\\
 ($M_\odot$) & \multicolumn{4}{c}{($\mbox{Gpc}
^{-3}\mbox{y}^{-1}$)} \\
 & $0.6M_\odot$ WD & $1.4M_\odot$ MWD/NS & $10M_\odot$ BH & $100M_\odot$ PopIII BH
\\  \hline
$10^{6.5\pm0.25}$ & 8.5 & 1.7 & 1.7 & $1.7\times 10^{-3}$\\
$10^{6.0\pm0.25}$ & 6 & 1.1 & 1.1 & $10^{-3}$ \\
$10^{5.5\pm0.25}$ & 3.5 & 0.7 & 0.7 & $7\times 10^{-4}$ \\
\hline
\end{tabular}
\caption{Estimated number of extreme mass ratio mergers. The rate of mergers per cubic $\mbox{Gpc}$ per year is given for three ranges in the mass of the supermassive black hole, $M_{\bullet}$, and four types of captured compact object.}
\label{ratetab}
\end{table}

\section{Detection of EMRIs}
\label{detect}
The amplitude of a typical EMRI event is below the level of noise fluctuations in the detector. Matched filtering allows detection of signals by building up signal power over many cycles of the waveform. However, the complexity of EMRI waveforms makes this procedure challenging. The inspiral waveform depends on 14 different parameters --- 2 angles and a distance defining the source location, 2 angles defining the orientation of the primary's spin, the mass and spin of the primary, the mass of the secondary, three adiabatic constants defining the geodesic that the secondary is on at a specified time $t_{0}$ (for instance the orbital periapse, eccentricity and inclination) and three dynamical constants that represent the phases of the secondary's motion at $t_{0}$ \cite{ghk}. In the final year of the inspiral, an EMRI waveform has $\sim 1 {\rm y} \times 3{\rm mHz}\sim 10^5$ cycles. Even assuming that only about eight of these fourteen parameters affect the phase evolution (see next paragraph) one would still estimate that $\sim(10^5)^8=10^{40}$ templates will be required to perform a fully coherent matched filter search for year long inspiral waveforms. The most generous extrapolation to 2013 still makes this far more than is computationally reasonable. An optimistic extrapolation of Moore's Law (doubling compute power every 1.5 to 2 years~\cite{itrs}) would yield typical commercial CPUs with around 50--100~Gflops when LISA flies, and around 50--100~Tflops for a cluster of a thousand such units.
%A typical high-end computer cluster today might have $1000$ nodes, each capable of $1.5$ Gflops. %Moore's Law predicts that a computer's capability in flops will double every one and a half to %two years (see for instance \cite{itrs}), so an equivalent cluster will be capable of between %$\sim 50$ Tflops and $\sim 150$ Tflops when LISA flies in about a decade. 
Computing the overlap with a template sampled at $\sim 0.03 {\rm Hz}$ 
uses $\sim 0.03$ flops, so this cluster could search $\sim 10^{15}$ 
templates. Even taking advantage of various tricks to reduce the template 
count (see below), the $10^{40}$ templates needed for a fully coherent 
search is well out of reach. Instead, we will have to use a hierarchical 
search, the first stage of which will be a coherent search of shorter 
segments of the data. The preceding parameter list did not include the spin of the secondary (3 additional parameters). This is a reasonable omission since it does not significantly affect the phase of the waveform over the duration of the short coherent segments.

The initial coherent search can be simplified by noting that the two sky position angles, the two source orientation angles and the azimuthal phase of the inspiraling body at $t_{0}$ are `extrinsic' --- these parameters do not change the intrinsic radiation of the source, but only how it projects onto the detector. This distinction makes it cheap to search over these variables \cite{bcv}. A pure quadrupole gravitational wave can be decomposed into a linear combination of five orthonormal waveforms $h_{i}({\bf \lambda}_{I};t)$ that depend on the intrinsic parameters, with amplitudes that depend on the extrinsic parameters. The optimal matched filter statistic
\begin{equation}
\fl \rho^{2}=\sum_{\alpha=I}^{II}\sum_{i=1}^{5}\,\left< h_{i}({\bf \lambda}_{I}),s_{\alpha}\right>^{2}, \,\,\,\, {\rm where} \,\,\,\, \left< a,b\right> = 4\,\Re\left[\int_{0}^{\infty} \frac{\tilde{a}^{*}(f)\,\tilde{b}(f)}{S_{b}(f)}\,\rmd f \right]
\label{rhosq}
\end{equation}
maximizes over possible values of the extrinsic parameters automatically. In \erf{rhosq}, $\tilde{a}$ denotes a Fourier transform, $z^*$ denotes complex conjugation and $S_{b}(f)$ is the power spectral density of the detector noise in the channel $b(t)$. The LISA output can be represented approximately by two synthetic Michelsons (denoted I and II) \cite{cutler98}, and the $s_{\alpha}$  are the signals constructed from these two channels. The expression \erf{rhosq} is an unconstrained maximum and so the best fit amplitudes need not correspond to physical values of the extrinsic parameters. The unconstrained maximum has the advantage of computational simplicity at the expense of  a slight increase in the false alarm probability for the search. The sky position angles are not strictly extrinsic, since the motion of LISA in its orbit introduces a sky position dependent Doppler modulation into the signal. However, on short ($\sim$ few week) timescales, this principally causes a linear frequency drift. This can be mimicked by redshifting the mass of the primary, which allows us to use \erf{rhosq} for the coherent search if the segments are not too long. True EMRI waveforms additionally have significant contributions from multipole moments other than the quadrupole. In this analysis we adopt the quadrupole search statistic \erf{rhosq} again for computational ease at the expense of some lost detection rate. Further computational savings may be gained by replacing one of the adiabatic parameters (say the periapse) with the time offset from when that parameter had a particular value, $r_{p}=r_{0}$. Time offsets can be searched cheaply using inverse fast Fourier transforms.

The signal-to-noise ratio (SNR) accumulated in a single short coherent segment is not sufficient for detection. The SNR is built up in the second stage of the search, by incoherently adding the power in the coherent segments to find inspirals that persist through the full data set. The phase angles of the motion in the radial and vertical directions (the Boyer-Lindquist $r$ and $\theta$ coordinates) vary on a dynamical timescale, and so a very high resolution in the other search parameters would be needed if we were to require consistency in these angles over the whole inspiral. For this reason, we maximize over these angles (using a template bank) before combining the coherent results. This is why we refer to this stage of the search as `incoherent'. The resulting statistic, $P_{k}$, on a given coherent segment is the maximum of $\rho^{2}$ over all possible values of these two dynamical phases. The azimuthal phase angle (Boyer-Lindquist $\phi$ coordinate) also varies on a dynamical timescale, but this is an extrinsic parameter and we maximize over it automatically by constructing the $\rho^2$ statistic. A given set of full inspiral parameters defines a trajectory through certain coherent segments at certain times. Summing $P_{k}$ along this trajectory gives the final search statistic. This scheme is a first-cut approach to a hierarchical search and more optimal approaches are currently being investigated.

\section{Estimating the LISA event rate}
\label{rate}
\subsection{Template waveforms}
Waveform templates are needed to scope out this approach to data analysis. The extreme mass ratio means that true waveforms can be well determined from perturbation theory. However, no gravitational waveforms have yet been generated using fully perturbative calculations, although this should be possible by the time LISA flies \cite{poisson03}. Waveforms from special classes of geodesic orbits have been computed \cite{gk2002,hughes2000} but these are neither generic enough nor sufficiently fast computationally to be useful. Instead, we have made use of two types of kludged waveforms. The `numerical kludge' waveforms are constructed by assuming the inspiraling body is always instantaneously on a geodesic of the spacetime. The geodesic parameters are evolved continuously using post-Newtonian radiation reaction expressions and an approximate quadrupole gravitational wave is then constructed from the resulting trajectory \cite{ghk}. The `analytic kludge' waveforms are constructed by adding post-Newtonian precessions and evolution to Peters and Mathews (Keplerian) waveforms \cite{AK03}. Neither set of waveforms has sufficient phase accuracy to be used for detection purposes, but the waveforms capture the main features of the inspiral and should therefore give reasonable estimates of template counts. The analytic and numerical kludge waveforms drift out of phase with each other over a timescale of a few hours, although this can be corrected somewhat by modifying the waveform parameters. However, the template counts obtained from the two approaches agree within a few tens of percent. Performing the analysis using these two independent methods in parallel has thus increased our confidence in the results. Comparisons to perturbative waveforms from geodesics at certain points in parameter space also indicate that the kludges perform extremely well.

\subsection{Length of coherent integration}
The maximum length of the coherent segments is set by computational limitations. Assuming that the data analysis will be performed on a 50 Teraflops computer cluster over $k$ years ($\sim$the mission lifetime), there are $\sim 1.5\,k\times10^{21}$ operations at our disposal. The number of operations required to search the mission data with a single coherent template is $10 \times 3f_{{\rm max}}\tau \log_{2}(f_{{\rm max}}\tau)$, where $\tau = k {\rm y} \sim 3 {\rm y}$ is the total length of the mission and $f_{{\rm max}} \sim 0.03 {\rm Hz}$ is the maximum frequency we try to resolve at the first stage. The factor $10$ is the cost to compute the $\rho^{2}$ statistic \erf{rhosq} and the rest is the cost of using FFT's to search over all possible time offsets. Equating the computational cost with the assumed computing power, we can cope with up to $\sim 10^{12}$ templates. This assumes that the coherent stage dominates the computational cost. Estimates of the cost of the incoherent stage indicate that this is a fair assumption.

To estimate the number of templates required for a coherent search of segments of duration $T$, we compute the standard metric on template space, $\Gamma_{ab}=\frac{1}{2}\left<\partial_{a}h|\partial_{b}h\right>$ \cite{bal96,owen}. Since the search over extrinsic parameters and the time offset have already been accounted for, we project these directions out of the metric to give the number of intrinsic templates
\begin{equation}
N_{{\rm int}} \approx \left(12 (1-{\cal A}) \right)^{-\frac{\hat{N}}{2}}\,\hat{N}^{\frac{\hat{N}}{2}}\,\int\sqrt{\gamma}\,\rmd\lambda^{1}\cdots\rmd\lambda^{\hat{N}},
\label{Ntemp}
\end{equation}
where $\hat{N}(=7)$ is the number of remaining parameters, $\gamma$ is 
the determinant of the $\hat{N}\times\hat{N}$ metric projected onto the 
intrinsic subspace and ${\cal A} < 1$ is the average match. The latter 
characterizes the coarseness of the template grid --- it is the average 
overlap that a randomly placed waveform has with the nearest template in 
the bank. There are seven not eight remaining parameters since the 
distance to the source is only an amplitude scaling and therefore does 
not require additional templates. The template count \erf{Ntemp} is 
averaged over values of the extrinsic parameters and the time offset. 
Assuming we are searching only for the last $k$ years of the inspiral, 
templates must be placed at approximately $N=k/(T/{\rm y})$ different 
values of the time offset to cover the whole inspiral, giving a total 
$N_{{\rm temp}}=N_{{\rm int}}\,N$. We have used Monte Carlo simulations 
to calculate the integral \erf{Ntemp}. This involved choosing points 
randomly distributed over the waveform parameter space and computing the 
corresponding determinant of the template metric, $\gamma$, at each 
point. We also looked at trends in the template density as the template parameters 
were slowly varied and found power 
law dependencies which were verified over large regions of parameter 
space by the Monte Carlo simulation. The power law fitting function was 
easily integrated to give the final template count \erf{Ntemp}. The 
template count was estimated using both types of kludged waveforms and the 
two results agreed to within a few tens of percent. Overall, our results 
suggested that, for an average match of $0.9$, the computational limit to 
the length of the coherent segments is $\sim 3~{\rm weeks}$. This first 
cut analysis assumed that the coherent templates were of equal length in 
time. In practice, it will be better to use a different division into 
coherent segments, for instance templates with equal numbers of wave 
cycles. Such refinements are being investigated.

\subsection{Threshold signal to noise}
\begin{figure}
\centerline{\includegraphics[keepaspectratio=true,height=6in,angle=0]{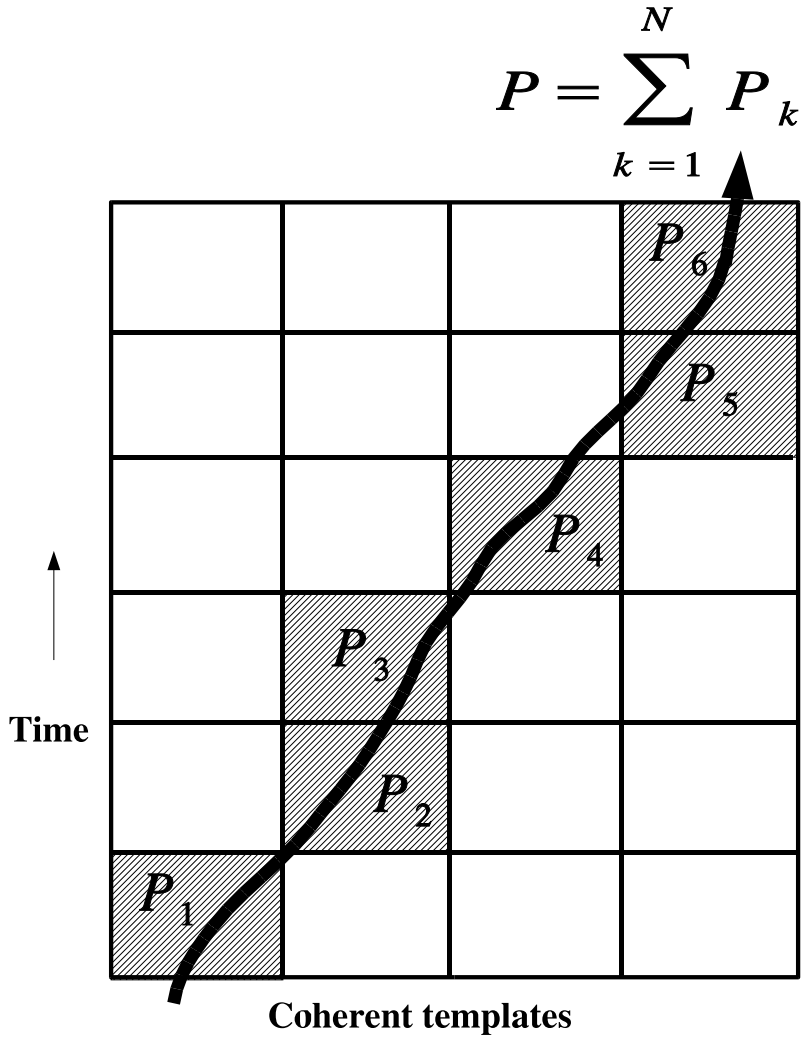}}
\caption{Illustration of the incoherent summation. A full inspiral corresponds to a sequence of the `best-overlapping' coherent templates at different times. Our final search statistic is obtained by summing the power along such trajectories. We maximize over the two phase angles before summation in order to reduce the computational cost of the incoherent stage.}
\label{threshfig}
\end{figure}

The final detection statistic in this scheme is $P=\sum_{k=1}^{N}P_{k}$ --- the sum of the coherent detection statistics, $P_{k}$, along a trajectory through certain coherent segments at certain times, which corresponds to a particular inspiral. This is illustrated pictorially in Figure~\ref{threshfig}. The coherent statistic $P_{k}$ is the maximum of the $\rho^{2}$ statistic along a slice in parameter space (corresponding to varying the two phase angles). In the presence of Gaussian noise, the $\rho^{2}$ statistic is distributed as a $\chi^{2}$ with $10$ degrees of freedom, but the distribution of the maximum, $P_{k}$, is analytically intractable. However, by performing Monte Carlo simulations over a range of signal parameters, we find that $P_{k}$ typically has mean $\mu_{k} \sim 18$ and standard deviation $\sigma_{k} \sim 4.5$. From the central limit theorem, we argue that the sum $P$ may be approximated by a normal distribution with mean $\mu=N\mu_{k}$ and standard deviation $\sigma=\sqrt{N}\sigma_{k}$. 

The detection threshold is characterized by the number of standard deviations by which $P$ must exceed its pure noise mean in order to be a robust detection. This is the threshold Z-score, $Z=(P-\mu)/\sigma$. We set the threshold to ensure the entire search has a false alarm probability $p_{f}$. Since the search involves a summation along many different possible trajectories, we need the false alarm rate for each trajectory to be less than $p_{f}/N_{{\rm traj}}$, where $N_{{\rm traj}}$ is the number of independent trajectories, i.e., the number of independent ways to combine the coherent segments into a full inspiral. For a fixed trajectory start time, we expect this number to exceed the $N_{{\rm int}}=N_{{\rm temp}}/N \sim10^{10}$ templates used for each time offset in the coherent search. During a three year observation there are additionally $\sim 3\,{\rm y}\times 1\,{\rm mHz} \sim 10^5$ independent start times, which means that for a $1\%
$ total false alarm rate, the false alarm rate per trajectory per time offset should be less than $10^{-17}$, giving a threshold Z-score of $8.8$. The number of independent trajectories can be more accurately estimated by computing the metric on the incoherent space in a similar way to the metric on the coherent space. Preliminary calculations suggest that the estimate of $\sim 10^{10} \times 10^5$ independent trajectories is reasonable. In fact, changing this by an order of magnitude does not significantly affect the threshold Z-score since we are well within the tail of the distribution. We assume a typical Z-score of $8$ in subsequent threshold calculations.

In the presence of a signal, the Z-score is increased. If a source has an intrinsic signal to noise ratio SNR$=\sqrt{\langle s,s \rangle}$ over the entire observation, the SNR in each of the $N$ coherent segments is approximately a factor of $1/\sqrt{N}$ less. The expected SNR in each segment is further reduced by the fact that we use a discrete bank of templates to detect the signal. We characterize the coarseness of the template bank by a match factor ${\cal M}$, which has contributions from discreteness in the coherent template bank (the average match ${\cal A}$), in the time sampling and in the incoherent template bank. We assume an overall match factor of ${\cal M}= 0.8$ is reasonable. A signal increases the expectation value of the $\rho^{2}$ statistic for the nearest template from $10$ to $10+({\cal M}/N)\,\langle s,s \rangle$. Assuming the maximization over the phase angles gives the correct values for the embedded signal, the corresponding $P_{k}$'s will have a similar value, and the sum $P$ will be $N$ times that value. Setting this equal to the threshold Z-score $Z_{{\rm thresh}}$, we find the threshold signal to noise for a detection
\begin{equation}
{\rm SNR}_{{\rm thresh}}=\sqrt{\frac{N}{{\cal M}} \,\left( \mu_{k} - 10 + Z_{{\rm thresh}}\,\frac{\sigma_{k}}{\sqrt{N}}\right)}
\label{threshsnr}
\end{equation}
This is the signal-to-noise ratio from the two synthetic Michelsons combined that a source must have in order to be detected under this scheme. The match factor ${\cal M}$ does not include a `fitting factor' (reduction in signal due to physical inaccuracies in the template model~\cite{owen}) since the actual search, when LISA flies, will use true perturbative waveforms as templates, for which the fitting factor should be close to one.

%If we were to use kludged waveforms as templates for the final search, we would have to reduce %the match factor ${\cal M}$ further to account for the `fitting factor' \cite{owen} of the %approximate waveforms, i.e., the loss in signal that arises because the kludged waveforms are not %an exact fit to the true signal waveforms. However, we envisage that the final search will be %performed using perturbative waveform templates, for which the fitting factor should be close to %one, so we can safely ignore this factor. Some allowance should be made in the template count %\erf{Ntemp} for the difference between the kludged waveforms and the realistic waveforms that %will be used in the final search, but we expect our current template counts to be fairly %representative and this correction to be small.

%There will be some difference between the number of templates we estimate using the kludged waveforms and the actual number required to search for true waveforms, %which will affect the maximum length of the coherent integration. We expect that the number of templates estimated here will be a good approximation to the true %template count.

\subsection{White dwarf background}
The population of close white dwarf binaries in the Milky Way radiate in the LISA band, generating a background that affects our ability to detect inspirals. At higher frequencies the binaries are sufficiently separated in frequency to be removed from the data. At lower frequencies we are theoretically limited by Shannon's Theorem, which gives the binary confusion noise used in \cite{AK03}. In practice, the best white dwarf subtraction algorithm that currently exists is the gCLEAN algorithm \cite{cornish03}. The residual noise remaining after applying the gCLEAN algorithm is somewhat higher than predicted by the Shannon limit, and it is possible to estimate the effective gCLEAN confusion noise limit from these residuals. We have computed signal-to-noise ratios using both confusion noise levels. We regard the Shannon limit as optimistic and the current gCLEAN limit as pessimistic in terms of how well we will be able to subtract binaries from actual LISA data.

\subsection{Estimated rates}
For a given source, the SNR scales with proper distance to the source as $D^{-1}$. The SNR threshold \erf{threshsnr} therefore translates into a maximum distance to which the source can be detected. If a given type of inspiral event is occurring at a rate ${\cal R}$ and has signal-to-noise ratio SNR at a fiducial distance $D$, then during an observation time $T$ we will detect $N_{det}$ events
\begin{equation}
N_{{\rm det}}=\frac{4\,\pi}{3}D^{3}{\cal R}T\left[\frac{{\rm SNR}}{{\rm SNR}_{{\rm thresh}}}\right]^{3}.
\label{Ndet}
\end{equation}
Using the capture rates in Table~\ref{ratetab}, we can estimate how many events of each type LISA will detect during its lifetime. We computed the appropriate SNR at the fiducial distance of $1{\rm Gpc}$ using the Synthetic LISA simulator \cite{vallis04} and numerical kludge waveforms. We fixed the masses of the primary and secondary at a typical value for each mass range, but averaged the SNR over possible orientations of the source and over two values of the eccentricity at plunge --- $e=0.25$ (corresponding to large initial periapse) and $e=0.4$ (small initial periapse). The averaging assumed the orientation angles are uniformly distributed and the eccentricities equally likely, but was appropriately weighted to account for the fact that certain parameter values allow the source to be detected to larger distances. The event rate estimates are given in Table~\ref{LISArate} for an `optimistic' case and a `pessimistic' case:
\begin{itemize}
\item `Optimistic' --- assumes a 5 year LISA lifetime, SNR's constructed from the optimal AET combination \cite{ptla02}, optimistic white dwarf background subtraction and 3 week coherent integrations (giving ${\rm SNR}_{{\rm thresh}}\sim 36$).
\item `Pessimistic' --- assumes a 3 year LISA lifetime, SNR's constructed from a single synthetic Michelson (X signal), pessimistic (gCLEAN) white dwarf background subtraction and 2 week coherent integrations (giving ${\rm SNR}_{{\rm thresh}}\sim 34$).
\end{itemize}
We do not fold uncertainties in the astrophysical event rate into this distinction, but use the rates in Table~\ref{ratetab} in both cases. Results are quoted for both the standard LISA design, with $5\times 10^6 {\rm km}$ arm length, and a `Short LISA' design, for which the arm lengths were reduced to $1.6 \times 10^6 {\rm km}$. Equation \erf{Ndet} assumes a flat Minkowski space volume/distance relation. For sources at redshift $z > 1$, this should not be trusted. Moreover, effects that have been ignored such as frequency redshifting and source evolution will become important. For cases with ${\rm SNR}(1{\rm Gpc}) \stackrel{>}{\scriptstyle\sim}120$ (so that $D_{{\rm max}} > 3.6 {\rm Gpc}=D_{p}(z=1)$), we quote a rough lower limit to the number of events with $z<1$, all of which LISA can detect --- $N_{{\rm det}} =V_{c}(z<1)\,{\cal R}T$, using comoving volume $V_{c}=199{\rm Gpc}^{3}$ (appropriate for a flat, $\Omega_{m}=0.27$, $H_{0}=65{\rm km}\,{\rm s}^{-1}\,{\rm Mpc}^{-1}$ cosmology).

\begin{table}
\begin{tabular}{cc|cc|cc} \hline
$M_\bullet$ & $m$ & \multicolumn{2}{|c|}{LISA} & \multicolumn{2}{c}{Short LISA} \\
&& Optimistic & Pessimistic & Optimistic & Pessimistic \\
\hline\hline
   300\,000 & 0.6 &     8 &   0.7 &    14 &     1 \\
   300\,000 &  10 &   700* &    89 &   902 &   115 \\
   300\,000 & 100 &     1*&     1*&     1*&     1*\\
\hline
1\,000\,000 & 0.6 &    94 &     9 &    80 &     7 \\
1\,000\,000 &  10 &  1100*&   660* &  1100*&   500 \\
1\,000\,000 & 100 &     1*&     1*&     1*&     1*\\
\hline
3\,000\,000 & 0.6 &    67 &     2 &    11 &   0.3 \\
3\,000\,000 &  10 &  1700*&   134 &   816 &    25 \\
3\,000\,000 & 100 &     2*&     1*&     2*&     1 \\ \hline
\end{tabular}
\caption{Estimated number of LISA EMRI detections, under the `optimistic' and `pessimistic' set of assumptions described in the text. Estimates are given for both the standard LISA design, and for `Short LISA'. Entries marked with a $^*$ are $z < 1$ lower limits since LISA can detect these sources to $z \gg 1$ and evolution is unknown.}
\label{LISArate}
\end{table}

\section{Discussion}
\label{discuss}
The estimates in Table~\ref{LISArate} indicate that even under the pessimistic assumptions, LISA should detect over a thousand EMRI events during its lifetime and there seems to be no particular advantage to using a shorter arm length. The biggest remaining uncertainty is in the astrophysical rates (Table~\ref{ratetab}). Using more conservative estimates, the detection of white dwarf inspirals becomes marginal, but the stellar mass black hole signals are still robust. The uncertainty arising from the use of kludged waveforms may be quantified by comparing to perturbative waveforms \cite{gk2002,hughes2000}. The template counts appear quite robust and the SNRs are not bad. We have compared SNRs computed from kludged waveforms to SNRs computed using the quadrupole piece of an adiabatic sequence of perturbative waveforms in the simplest case of circular equatorial inspiral. For such inspirals, the last year SNRs agree within $\sim20\%
$ for all spins, but the kludge SNRs are larger. This could lead to a factor $\sim2$ overestimate of the detection rates. However, including all multipoles, the total SNRs of the perturbative
waveforms are larger than those of the pure quadrupole kludge waveforms, 
since the higher multipoles are generally at frequencies more accessible to LISA.  
A modified search therefore might recover these events. This effect is 
expected to be even stronger for eccentric orbits although the 
uncertainty in the rates should still only be a factor of a few, but 
perturbative eccentric inspiral waveforms were not available for 
comparison. 
%The disagreement may be somewhat larger for non-equatorial and eccentric inspirals, but the %necessary perturbative inspiral waveforms are not currently available to compare to our results. %However, the comparison in the circular equatorial case and comparisons to perturbative waveforms %from geodesics \cite{gk2002} suggest that the resulting uncertainty in the rate should only be a %factor of a few. 

An issue that has not yet been properly addressed is self-confusion --- there is a background of gravitational radiation from all the unresolved EMRIs in the Universe from which the louder signals must be extracted. This is an additional noise source which is now being estimated. We have also assumed that it is possible to remove the resolved white dwarf binaries from the LISA data stream before performing the EMRI search. Methods like gCLEAN in principle will remove the EMRI signals as well as the white dwarf binaries. It may therefore be necessary to do the searches for EMRIs and binaries simultaneously, which would change the computational requirements.

The approach to EMRI data analysis outlined here is a first cut. Under this scheme a source requires a SNR $\sim30$ to be detected. If we had infinite computational power and could perform a fully coherent search, the detection SNR would still be $\sim 15$, for the same false alarm probability. We therefore lose about a factor of two in reach by using a hierarchical search. With a more careful division into coherent segments or a different incoherent summation, we may recover some of these lost events. The fact that even this first cut scheme predicts a thousand inspiral detections suggests that LISA will detect many EMRI events despite the remaining uncertainties.

\ack This work would not have been possible without the guidance and encouragement of Kip Thorne, who repeatedly cracked the whip to make sure the project stayed on schedule. We also thank Kostas Glampedakis, Scott Hughes and Daniel Kennefick for the use of their kludge waveform code and for many useful discussions at EMRI telecons. This work was supported in part by NASA grants NAG5-12834 (CC, JG) and NAG5-10707 (JG, ESP), LISA contract number PO 1217163 (SL, TC), a Marie Curie Fellowship of the European Community program IHP-MCIF-99-1 under contract number HPMF-CT-2000-00851 (LB), NSF Grant NSF-PHY-0140326 (`Kudu') (LB), a grant from NASA-URC-Brownsville (LB) and by the LISA Mission Science office at JPL under contract with NASA (MV).


\section*{References}
\begin{thebibliography}{30}
\bibitem{LISAppa} Danzmann K \etal 1998 {\it LISA - Laser Interferometer Space Antenna, Pre-Phase A Report}, Max-Planck-Institut f\"{u}r Quantenoptic Report MPQ 233 
\bibitem{ryan97} Ryan F D 1997 \PR D {\bf 56} 1845
\bibitem{gk2002} Glampedakis K and Kennefick D 2002 \PR D {\bf 66} 044002
\bibitem{AK03} Barack L and Cutler C 2004 \PR D {\bf 69} 082005
\bibitem{collins04} Collins N and Hughes S 2004 grqc/0402063
\bibitem{LIST03} Barack L, Creighton T, Cutler C, Gair J, Larson S, Phinney E S, Thorne K S and Vallisneri M 2003 {\it `Estimates of detection rates for LISA capture sources'} report prepared by working group 1 for LIST meeting December 2003, available at http://www.tapir.caltech.edu/$\sim$listwg1/EMRI/LISTEMRIreport.pdf
\bibitem{gamm04} Gammie C F, Shapiro S L and McKinney J C 2004 \ApJ {\bf 602} 312
\bibitem{mckinn04} McKinney J C and Gammie C F 2004 \ApJ {\it in press} astro-ph/0404512
\bibitem{mf01} Merritt D and Ferrarese L 2001 \ApJ {\bf 547} 140
\bibitem{trem02} Tremaine S \etal 2002 \ApJ {\bf 574} 740
\bibitem{aller02} Aller M C and Richstone D 2002 \AJ {\bf 124} 3035
\bibitem{Scobs} Gebhardt K \etal 2001 \AJ {\bf 122} 2469; Filippenko A V and Ho L C 2003 \ApJ{\em Lett.} {\bf 588} L13
\bibitem{freitag01} Freitag M 2001 \CQG {\bf 18} 4033
\bibitem{madau01} Madau P and Rees M J 2001 \ApJ{\it Lett.} {\bf 551} L27
\bibitem{sig97} Sigurdsson S and Rees M J 1997 \MNRAS {\bf 284} 318
\bibitem{binney88} Binney J and Tremaine S 1988 {\it Galactic Dynamics} (Princeton University Press)
\bibitem{sheth03} Sheth R K, Bernardi M, Schechter P \etal 2003 \ApJ {\bf 594} 255
\bibitem{hils95} Hils D and Bender P 1995 \ApJ{\it Lett.} {\bf 445} L7
\bibitem{freitag03} Freitag M 2003 \ApJ{\it Lett.} {\bf 583} L21
\bibitem{ghk} Glampedakis K, Hughes S and Kennefick D 2002 \PR D {\bf 66} 064005
\bibitem{itrs} `{\em International Technology Roadmap for Semiconductors}' 2003 \newline http://public.itrs.net/Files/2003ITRS/Home2003.htm
\bibitem{bcv} Buonanno A, Chen Y and Vallisneri M 2003 \PR D {\bf 67} 024016
\bibitem{cutler98} Cutler C 1998 \PR D {\bf 57} 7089
\bibitem{poisson03} Poisson E 2003 {\it Living Reviews in Relativity}, submitted gr-qc/0306052
\bibitem{hughes2000} Hughes S 2000 \PR D {\bf 61} 084004; Hughes S 2001 \PR D {\bf 64} 064004
\bibitem{bal96} Balasubramanian R, Sathyaprakash B S and Dhurandhar S V 1996 \PR D {\bf 53} 3033; Erratum-ibid. 1996 \PR D {\bf 54} 1860
\bibitem{owen} Owen B J 1996 \PR D {\bf 53} 6749
\bibitem{cornish03} Cornish N J and Larson S L 2003 \PR D {\bf 67} 103001
\bibitem{vallis04} Vallisneri M 2004 {\it Synthetic LISA: Simulating Time Delay Interferometry in a Model LISA}, in preparation
\bibitem{ptla02} Prince T A, Tinto M, Larson S L and Armstrong J W 2002 \PR D {\bf 66} 122002
\end{thebibliography}
\end{document}